\documentclass[12pt]{article}
\usepackage{amsmath,amssymb}
\usepackage{graphics}
\usepackage[dvipdfm]{graphicx}
\usepackage{color}
\usepackage{url}
\makeatletter
\def\alt{\mathrel{\mathpalette\gl@align<}}
\def\agt{\mathrel{\mathpalette\gl@align>}}
\def\gl@align#1#2{\lower.6ex\vbox{\baselineskip\z@skip\lineskip\z@
\ialign{$\m@th#1\hfil##\hfil$\crcr#2\crcr\sim\crcr}}}
\makeatother

\begin{document}
\begin{titlepage}

\begin{center}

{\Large\bf 
Relation between proton decay and PMNS phase \\
in the minimal SUSY $SO(10)$ GUT
} 
\lineskip .75em
\vskip 1.5cm

\normalsize
{\large Takeshi Fukuyama},$^1$ 
 {\large Koji Ichikawa},$^2$
and
 {\large Yukihiro Mimura}$\,^3$

\vspace{1cm}

$^1${\it  Research Center for Nuclear Physics (RCNP), Osaka University, \\ 
Ibaraki, Osaka, 567-0047, Japan} \\

$^2${\it  University of Tokyo, \\ 
Kavli Institute for the Physics and Mathematics of the Universe (KIPMU) \\
5-1-5 Kashiwa-no-Ha, Kashiwa Shi, Chiba 277-8568, Japan} \\

$^3${\it Department of Physics, 
 National Taiwan University, %No. 1, Section 4, Roosevelt Rd.,
 Taipei, 10617, Taiwan} \\

\vspace*{10mm}

{\bf Abstract}\\[5mm]
{\parbox{13cm}{\hspace{5mm}
%
%%%%%%%%%%%%%%%%%%%%%%%%%%%%%%%%%%%%%%%%%%%%%%%%%%%%%%%%%%%%%%%%
%             ABSTRACT                                         %
%%%%%%%%%%%%%%%%%%%%%%%%%%%%%%%%%%%%%%%%%%%%%%%%%%%%%%%%%%%%%%%%

Proton decay is one of the most important predictions of the grand unified theory (GUT).
In the supersymmetric (SUSY) GUT, proton decays via the dimension-five operators need to be suppressed.
In the $SO(10)$ model where ${\bf 10}+\overline{\bf 126}$ Higgs fields couple to fermions,
neutrino oscillation parameters including the CP-violating 
Pontecorvo-Maki-Nakagawa-Sakata (PMNS) phase 
can be related to the Yukawa couplings to generate the dimension-five operators in the unified framework.
We show how the suppressed proton decay depends on the PMNS phase,
and stress the importance of the precise measurements of the PMNS phase
as well as the neutrino 23-mixing angle.
These become especially important 
if the SUSY particles are found around less than a few TeV at LHC
and proton decays are observed at Hyper-Kamiokande and DUNE experiments
in the near future.

\vspace*{7mm}

%{\bf PACS.} 12.10.-g, 12.10.Dm, 12.10.Kt
}}
\end{center}

\end{titlepage}

%%%%%%%%%%%%%%%%%%%%%%%%%%%%%%%%%%%%%%%%%%%%%%%%%%%%%%%%%%%%
%                         Introduction                     %
%%%%%%%%%%%%%%%%%%%%%%%%%%%%%%%%%%%%%%%%%%%%%%%%%%%%%%%%%%%%

\baselineskip=16pt

\section{Introduction}

The measurement of the parameters in the neutrino oscillations is one of the recent major progress in particle physics.
Neutrinos must be very light compared to the other fermions in the standard model (SM).
The lightness of neutrinos can be explained by the seesaw mechanism
\cite{seesaw,Schechter:1980gr},
which can open a window to the ultra high scale physics.
In that sense, the precise measurements of the neutrino mixing angles and 
the PMNS phase \cite{An:2012eh,Wilking:2013vza,GonzalezGarcia:2012sz,Capozzi:2013csa}
are important not only for the scientific interests to archive the accurate values, but also to probe the physics beyond the SM.
In fact, their precise measurements are very important to determine the detail structure of the neutrino mass matrix.

The quantum numbers of quarks and leptons are disparate each other,
and it has been discussed whether quarks and leptons can be described in a unified framework.
In the $SO(10)$ GUT, the entire fermion species (including a right-handed neutrino) 
can be unified in one spinor multiplet in each generation,
and the description of fermion masses and mixings 
can be developed in a unified framework.
There are several ways to construct a model in the $SO(10)$ GUT.
Among them, the model with $\overline{\bf 126}$ Higgs representation has an attractive feature 
that the fermion coupling to $\overline{\bf 126}$ can generate both left- and right-handed
Majorana neutrino masses as well as a part of the Dirac masses of the fermions.
In that model, the smallness of quark mixing angles and two large neutrino mixings can be
naturally explained if the $\bf10$-dimensional Higgs coupling generates the third generation Yukawa couplings
dominantly \cite{Dutta:2009ij}.

Proton decay is one of the most important predictions of GUTs.
SUSY can solve the hierarchy problem between the weak scale and the GUT scale.
The SUSY GUTs, however, suffer from constraints of dimension-five proton decay operators \cite{Weinberg:1981wj},
which can be generated by the Yukawa interaction via color triplet Higgs exchange:
More than 10 years ago, one of the authors and his collaborators discussed 
how the proton decay amplitudes can be suppressed by the specific Yukawa structure,
which can be connected to the parameters in the neutrino oscillations,  
and claimed that the neutrino 13-mixing angle $\theta_{13}$ cannot be zero and $\theta_{13} \sim 0.1$ under the condition \cite{Dutta:2004zh}.
In these arguments, 
it was assumed that light neutrino mass matrix is dominated by the type II contribution from the left-handed Majorana neutrino mass.
In Ref.\cite{Fukuyama:2015kra}, we describe the sum rule of the fermion mass matrices in the form that 
the coupling matrices are given by the parameters in the lepton sector. 
Using the algebraic expression, we argue that 
the similar Yukawa structure can be applied even if 
the type I seesaw contribution is incorporated to the neutrino mass matrix in the case where
%the right-handed neutrino mass is large (but not decouple), namely 
the vacuum expectation value (VEV) of $\overline{\bf 126}$, $v_R$,
%to reduce the rank of the $SO(10)$ symmetry 
is large.

In this paper, 
we describe the Yukawa structure which is suitable to suppress the proton decay amplitude in the $SO(10)$ model,
and study how the Yukawa structure depends on the neutrino oscillation parameters.
In order to explain the proton decay suppression, 
we adopt the minimal $SO(10)$ model, in which the ${\bf 10}+\overline{\bf 126}$ Higgs fields couple to 
fermions \cite{Babu:1992ia,Matsuda:2000zp,Fukuyama:2002ch,clark,Aulakh:2004hm,Bajc:2002iw,Babu:2005ia,Bertolini:2006pe}.
We use the $\chi^2$ fits in the minimal $SO(10)$ model,
obtained in Ref.\cite{Fukuyama:2015kra} to show the consequence numerically.
In the case where $v_R \agt 10^{16}$ GeV,
the suppressed proton decay width depends on the parameters, 
and we show the numerical calculation in terms of 
the neutrino 23-mixing angle and the PMNS phase.
The accurate measurements of the oscillation parameters
 are important if the dimension-five proton decays are suppressed but it will be observed in the near future.

\section{The importance of the PMNS phase to the structure of the neutrino mass matrix}

CP violation of the neutrino oscillations is important to learn the nature of the fundamental symmetry.
Not only because of the fundamental curiosity, but also to understand the structure of the neutrino mass matrix
the precise measurement of the PMNS phase is necessary.

In the quark sector, the mixing angles are all small, and the hierarchical structure of the quark mass matrices
can be simply obtained.
The possible exception is the (1,1) element of the mass matrix.
When the mass matrix is reconstructed from various observations, there may be a cancellation in the (1,1) element.
It is well-known that there is an empirical relation between the quark mass ratio and the Cabibbo angle
\begin{equation}
\sin\theta_{C} \simeq \sqrt{\frac{m_d}{m_s}},
\end{equation}
which may be related to the smallness of the (1,1) element (compared to down quark mass) of the down-type quark mass matrix
in the basis where up-type quark mass matrix is diagonal.
In the lepton sector, on the other hands, the neutrino mixing angles
are large and the hierarchical structure is not obvious yet.
As in the quark sector, there may be a cancellation when the neutrino mass matrix is reconstructed from observations.

The neutrino mass matrix (in the basis where the charged lepton mass matrix is diagonal) is expressed as
\begin{equation}
{\cal M}_\nu = U {\rm diag}. (m_1,m_2,m_3) U^T,
\end{equation}
where $U$ is the unitary mixing matrix determined by three neutrino mixing angles $\theta_{12}, \theta_{23}, \theta_{13}$
and a phase $\delta_{\rm PMNS}$.
Using a convention by Particle Data Group \cite{Agashe:2014kda}, 
we obtain the (1,1) and (1,2) elements as
\begin{eqnarray}
({\cal M}_\nu)_{11}
&=& 
(m_1 \cos^2\theta_{12} + m_2 \sin^2\theta_{12}) \cos^2\theta_{13}
+ e^{-2i\delta_{\rm PMNS}} m_3  \sin^2\theta_{13}, \\
({\cal M}_\nu)_{12}
&=& 
\cos\theta_{13}
\left[(m_2-m_1)\cos\theta_{12}\sin\theta_{12}\cos\theta_{23} \right. \nonumber \\
&& 
\left.
+ e^{-i\delta_{\rm PMNS}} \sin\theta_{13} \sin\theta_{23} ( m_3
- e^{2i\delta_{\rm PMNS}}(m_1 \cos^2\theta_{12} + m_2 \sin^2\theta_{12}) )\right].
\end{eqnarray}
In this expression, mass eigenvalues $m_1, m_2, m_3$ are complex,
and one of three can be made to be real.
Because $\sin\theta_{13}$ and $\sqrt{{\Delta m^2_{\rm sol}}/{\Delta m^2_{\rm atm}}}$ are the same order,
it is possible that both (1,1) and (1,2) elements are cancelled\footnote{
We note that we here claim that the cancellation can happen among the mixing angles and the mass ratio
when the mass matrix is reconstructed from the observations.
It does not call a fine-tuning issue since 
the mixing angles and mass eigenstates are obtained from the mass matrix, and the structure of the matrix is more essential.
There may be a fundamental symmetrical reason that some of the elements are smaller than the naive size of them,
but in this paper, we take a stance only to discuss what can be expected if the elements are suppressed.
}
in the normal hierarchy (NH) case, $\sqrt{{\Delta m^2_{\rm sol}}/{\Delta m^2_{\rm atm}}} \sim m_2/m_3$.
We remark that the PMNS phase has to be chosen if those elements vanishes contrary to the quark sector,
and thus the accurate measurement of the PMNS phase is quite important to determine the 
structure of the neutrino mass matrix.

Let us suppose that both (1,1) elements and (1,2) elements are completely zero \cite{Fritzsch:2011qv,Geng:2015oga}
\begin{equation}
({\cal M}_\nu)_{11} = ({\cal M}_\nu)_{12} =0,
\end{equation}
to see what condition is needed to suppress those elements.
There are four equations for real degree of freedom.
We obtain a relation among the phase $\delta_{\rm PMNS}$ and mixing angles $\theta_{12},\theta_{23},\theta_{13}$
and the ratio of the mass squared differences $\Delta m^2_{12}/\Delta m^2_{23}$,
as well as $|m_1|$ and  two Majorana phases of the mass eigenvalues.
For the point of view that the constraint can be verified from the neutrino oscillations,
the relation among the phase, mixing angles and masses is the most important.
We obtain \cite{Haba:2011pe,Dutta:2013bvf}
\begin{equation}
\cos\delta_{\rm PMNS} = 
\frac{ \frac{\Delta m_{\rm sol}^2}{\Delta m_{\rm atm}^2} \cos2\theta_{13}\sin^22\theta_{12}   - 
4\sin^2\theta_{13} \left(\frac{\Delta m_{\rm sol}^2}{\Delta m_{\rm atm}^2} \cos^4\theta_{12}+\cos2\theta_{12}\right)
\tan^2\theta_{23}}{4\sin^3\theta_{13} \left(1+  \frac{\Delta m_{\rm sol}^2}{\Delta m_{\rm atm}^2} \cos^2\theta_{12}\right)\sin2\theta_{12}\tan\theta_{23}}.
\label{relation-PMNS}
\end{equation}
The atmospheric neutrino oscillations constrains $\sin2\theta_{23}$ (up to matter effects),
and thus $\tan\theta_{23}$ still has a large ambiguity.
We plot the relation between $\delta_{\rm PMNS}$ and $\theta_{23}$ in Fig.1
using the experimental measurements for the other parameters.
The experimental global fit results are overlapped in the figure.
One can see that the current experimental results are consistent with the conditions
of the suppressed (1,1) and (1,2) elements.

About a decade ago, one of the authors (Y.M.) and his collaborators argued 
that the (1,1) and (1,2) elements should be small 
and the 13-mixing angle is predicted to be a non-zero value and about 0.1 
if the proton decay amplitudes are suppressed in a renormalizable $SO(10)$ model \cite{Dutta:2004zh}.
Now, the 13-mixing angle is accurately measured,
and the key ingredient of the suppression shifts to the accurate values of the PMNS phase and 23-mixing angle.
While the PMNS phase lie around $\delta_{\rm PMNS}/\pi \sim -0.5$
in the global analysis, the 23-mixing angle still have ambiguity depending on 
which experiments have a major weight.
In fact, the global analysis in Ref.\cite{GonzalezGarcia:2012sz,Capozzi:2013csa} shows that the best fit lies at $\theta_{23}< \pi/4$,
however, it seems that the current best fit without the short-baseline reactor experiments lies at $\theta_{23}> \pi/4$.
If the 13-mixing angle is $1\sigma$ smaller than the current center value,
the best fit would shift to $\theta_{23}>\pi/4$
even in the global analysis.
More experimental data are needed.

\begin{figure}[t]
\begin{center}
\includegraphics[width=0.5\textwidth]{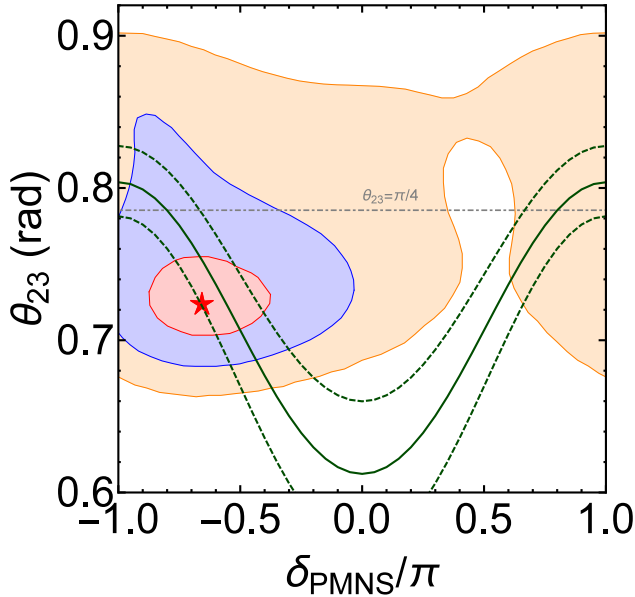}
\caption{
The plot from the relation in Eq.(\ref{relation-PMNS}).
Along the green line, the (1,1) and (1,2) elements can become zero in the neutrino mass matrix
in the basis where the charged-lepton mass matrix is diagonal.
The dotted lines are drawn using the 3$\sigma$ range of the mixing parameters.
We also overlap the current 1$\sigma$ (red), 2$\sigma$ (blue), 3$\sigma$ (orange) region of the global analysis 
in Ref.\cite{Capozzi:2013csa}.
The star symbol shows the current best fit of the global analysis.
}
\end{center}
\label{run1}
\end{figure}

\section{The proton decay suppression in the minimal $SO(10)$ model}

In the $SO(10)$ models,
the structure of the neutrino mass matrix can be related to the Yukawa couplings of the charged fermions. 
Therefore, proton decay amplitudes can be related to the neutrino mixing parameters.
In this section, we explain how the smallness of the (1,1) and (1,2) elements discussed in the previous section
is connected to the proton decay suppression using the minimal $SO(10)$ model.

The minimal $SO(10)$ model \cite{Babu:1992ia}
is defined as that in which the Yukawa interaction is minimal,
namely, only $H: {\bf 10}$ and $\bar\Delta: \overline{\bf 126}$ Higgs representations couple
to the fermions $\psi: {\bf 16}$ by renormalizable interaction:
\begin{equation}
W_Y= \frac12  {\bf h}_{ij} \psi_i \psi_j H + \frac12  {\bf f}_{ij} \psi_i \psi_j \bar \Delta.
\end{equation}
Due to the $SO(10)$ algebra, the coupling matrices are symmetric, ${\bf h}_{ij} = {\bf h}_{ji}$ and ${\bf f}_{ij} = {\bf f}_{ji}$.
The minimality assumption is often addressed even to the Higgs sector to break the $SO(10)$ symmetry,
but we impose the minimality just on the Yukawa interactions in this paper.

The Yukawa coupling (after $SO(10)$ symmetry is broken down to SM) is given by linear combinations of 
the $\bf h$ and $\bf f$ matrices,
%\begin{eqnarray}
%&& Y_u = h + r_2 f, \nonumber \\
%&& Y_d = r_1 (h + f), \nonumber \\
%&& Y_e = r_1 (h - 3 f), \nonumber \\
%&& Y_\nu = h- 3r_2 f,
%\end{eqnarray}
%
and the fermion mass matrices are obtained as
\begin{eqnarray}
M_u &=& (h+r_2 f) v_u, \nonumber\\
M_d &=& r(h+f) v_u, \nonumber\\
M_e &=& r(h-3f) v_u, \nonumber\\
M_\nu^D &=& (h-3r_2 f) v_u,
\end{eqnarray}
where $r$ and $r_2$ depend on the Higgs mixing (doublet Higgs mixing in {\bf 10} and $\overline{\bf 126}$),
and 
$h$ and $f$ are original Yukawa matrices $\bf h$ and $\bf f$ multiplied by Higgs mixings,
\begin{eqnarray}
h = V_{11} {\bf h}, \qquad
f = \frac{U_{12}}{\sqrt3 r_1} {\bf f}, 
\qquad
r = r_1 \frac{v_d}{v_u},
\qquad
r_1 = \frac{U_{11}}{V_{11}},
\qquad
r_2 = r_1 \frac{V_{13}}{U_{12}},
\end{eqnarray}
where
the unitary matrices $U$ and $V$ are the diagonalizing matrices of the doublet Higgs matrix $M_{\rm doublet}$:
$U M_{\rm doublet} V^T$ is diagonal,
and $v_u$ and $v_d$ are
the VEVs of up- and down-type Higgs fields.
The detail of the convention is found in Ref.\cite{Dutta:2004zh}.
We note that 
$r\sim m_b/m_t$ if $f_{33}$ is small,
and
$\tan\beta =v_u/v_d$ is roughly related to $r_1$, 
\begin{equation}
\tan\beta \sim r_1 \frac{m_t}{m_b}. 
\end{equation}

The right-handed Majorana neutrino mass matrix is obtained
as
\begin{equation}
M_R = \sqrt2 {\bf f} v_R,
\end{equation}
where $v_R$ is a VEV of $\overline{\bf 126}$.
Practically, we denote
\begin{equation}
M_R = c_R v_R f.
\end{equation}
The seesaw neutrino mass matrix can be written as \cite{seesaw, Schechter:1980gr} 
\begin{equation}
{\cal M}_\nu = M_L - M_\nu^D M_R^{-1} (M_\nu^D)^T,
\end{equation}
where
$M_L$ is the left-handed neutrino Majorana mass which comes from $SU(2)_L$ triplet coupling \cite{Cheng:1980qt},
$\ell \ell \Delta_L$.
In the $SO(10)$ model,
the $\overline{\bf 126}$ Higgs also includes the $SU(2)_L$ triplet $\Delta_L$
and the Yukawa coupling generates both $M_L$ and $M_R$.
Therefore, $M_L$ is also proportional to the coupling matrix $f$ and we denote 
\begin{equation}
M_L = c_L v_L f.
\end{equation}
In the $SO(10)$ model, 
if there is {\bf 210} or ${\bf 54}$ Higgs representation,
the VEV of $\Delta_L$, $\langle \Delta_L \rangle = v_L$, can be obtained as
$v_{\rm weak}^2/M_\Delta$, where $M_\Delta$ is the mass of the $SU(2)_L$ triplet.

The dimension-five operator is given in the form of superpotential as
\begin{equation}
-W_5 = \frac12 C_L^{ijkl} q_k q_l q_i \ell_j + C_R^{ijkl} e^c_k u^c_l u^c_i d^c_j.
\end{equation}
In the minimal $SO(10)$ model,
the dimension-five operators 
generated by the exchange of
the colored triplets, $({\bf 3},{\bf 1},-1/3)+(\bar{\bf 3},{\bf 1},1/3)$,
can be written as 
\begin{eqnarray}
C_L^{ijkl} &=& (M_T^{-1})_{11} {\bf h}_{ij} {\bf h}_{kl} + (M_T^{-1})_{12} {\bf f}_{ij} {\bf h}_{kl} 
+ (M_T^{-1})_{31} {\bf h}_{ij} {\bf f}_{kl} + (M_T^{-1})_{32} {\bf f}_{ij} {\bf f}_{kl} \label{cl1}\\
&=& \sum_a \frac1{M_{T_a}} (X_{a1} {\bf h} + X_{a2} {\bf f})_{ij} (Y_{a1} {\bf h} + Y_{a3} {\bf f})_{kl}, \label{cl2}\\
C_R^{ijkl} &=& (M_T^{-1})_{11} {\bf h}_{ij} {\bf h}_{kl} - (M_T^{-1})_{12} {\bf f}_{ij} {\bf h}_{kl} \label{cr1} \\
&&- ((M_T^{-1})_{31}-\sqrt2 (M_T^{-1})_{41}) {\bf h}_{ij} {\bf f}_{kl} 
+ ((M_T^{-1})_{32}-\sqrt2 (M_T^{-1})_{42}) {\bf f}_{ij} {\bf f}_{kl}  \nonumber \\
&=& \sum_a \frac1{M_{T_a}} (X_{a1} {\bf h} - X_{a2} {\bf f})_{ij} (Y_{a1} {\bf h} - (Y_{a3} - \sqrt2 Y_{a4}) {\bf f})_{kl}, \label{cr2}
\end{eqnarray}
where
$X$ and $Y$ are the diagonalization unitary matrices of the colored triplet mass matrix $M_T$:
\begin{equation}
X M_T Y^T = {\rm diag.} (M_{T_1}, M_{T_2}, \cdots, M_{T_5}).
\end{equation}
We denote that $M_{T_1}$ is the lightest triplet Higgs mass.
See Ref.\cite{Dutta:2004zh} for more details.

The dimension-five proton decay amplitude depends on 
(1) the masses of SUSY particles, (2) the triplet Higgs mass, and (3)
the structure of the Yukawa couplings, for the model parameters.
Among them, we here discuss the Yukawa structure to suppress the amplitudes.
People often consider the contribution only from the left-handed operator $C_L$,
since the contribution from the right-handed operator $C_R$ is smaller than $C_L$.
However, for the current experimental bounds, there must be a cancellation among the left-handed contributions,
and the contribution from $C_R$ itself can exceeds the bounds.
In fact, it turns out that the minimal $SU(5)$ model conflicts with the bounds 
due to the contribution from $C_R$ \cite{Goto:1998qg,Murayama:2001ur}.
In $SO(10)$ model, on the other hands,
there are additional freedoms to cancel the amplitudes in the triplet Higgs mixings,
and the amplitudes can be canceled, which is performed in Ref.\cite{Goh:2003nv}.
However, 
the signs of the ${\bf f}_{ij} {\bf h}_{kl}$ coefficients are opposite between $C_L$ and $C_R$
due to the algebraic reason that the triplet components in $H$ and $\bar \Delta$ have opposite ``D-parities".
This fact makes the simultaneous cancellation of both $C_L$ and $C_R$ unnatural \cite{Dutta:2004zh}.
More concretely, for the left-handed operator, the triplet Higgs mixing in Eq.(\ref{cl1}) can be chosen
to obtain $C_L^{ijkl} \sim (Y_u)_{ij} (Y_u + b f)_{kl}/M_{T_1}$ very roughly so that 
the amplitudes from $C_L^{112l}$ and $C_L^{121l}$ are suppressed.
Note that the $ij$ part of the operator in the minimal $SU(5)$ model (without a Higgs mixing) is given as $Y_d$ (or $Y_e$).
Even if the $i=j=1$ contribution is the size of the down quark Yukawa coupling, it can exceed the current experimental bound.
Then, the right-handed operator given in Eq.(\ref{cr1}) can be roughly written as
\begin{equation}
C_R^{ijkl} \sim (Y_u+ a f)_{ij} (Y_u+c f)_{kl}/M_{T_1},
\end{equation}
and $a \simeq 2 X_{12}$ cannot be zero as long as the charged fermion masses and mixing are fit by the
$h$ and $f$ matrices.
The right-handed operator generates the four-fermi proton decay operator by Higgsino-dressing,
$u_R d_R s_L \nu_\tau$ and $u_R s_R d_L \nu_\tau$,
and thus, the dominant contribution can be obtained as $C_R^{1133} y_t y_\tau V_{ts} + C_R^{1233} y_t y_\tau V_{td}$.
As a consequence, if the elements $f_{11}$ and $f_{12}$ are small (in the basis where the quark mass matrices are diagonal),
both the left- and right-handed contributions can be suppressed simultaneously.

We have learned that the proton decay amplitude can be suppressed if both $f_{11}$ and $f_{12}$ are small
for the given SUSY spectrum and triplet Higgs mass. 
The next issue is whether the smallness of the elements are realized both in the $f$ matrix and neutrino mass matrix.
The simplest realization is the case where the $SU(2)_L$ triplet contribution dominates the neutrino mass matrix:
${\cal M}_\nu = M_L$, and the neutrino mass matrix is proportional to $f$.
This can happen if the right-handed neutrino decouples and the type I contribution is negligible.
The fermion masses and mixings can be fit in the model that the fermions couples to ${\bf 10}+\overline{\bf 126}+ {\bf 120}$ Higgs
representations and the predictions can be obtained from the proton decay suppression \cite{Dutta:2004zh,Dutta:2013bvf}.
In the minimal model, on the other hand,
the triplet dominant neutrino mass does not provide a good fit,
and the type I contribution is necessary \cite{Babu:2005ia,Bertolini:2006pe}.
In general, if the type I term contributes, the $f_{11}$ and $f_{12}$ components are not correlated 
to the elements in the neutrino mass matrix.
However, we can show that 
 the coupling $f$ can be written as a linear combination of 
${\cal M}_\nu$, $M_e$ and ${\cal M}_\nu M_e^{-1} {\cal M}_\nu$ algebraically,
and the smallness of those elements can be correlated
in the case where $v_R$ is large ($c_R v_R \agt 10^{16}$ GeV) \cite{Fukuyama:2015kra}.
In the limit of $v_R \to \infty$, we obtain
\begin{equation}
c_L v_L f = \frac12 M_e^{1/2} (K+\sqrt{K^2}) M_e^{1/2},
\end{equation}
where
\begin{equation}
K = M_e^{-1/2} {\cal M}_\nu M_e^{-1/2}.
\end{equation}
The square root matrix $\sqrt{A}$ is defined as $(\sqrt{A})^2 = A$,
and there are $2^3 = 8$-fold branches in a $3\times 3$ matrix. 
As one can see, in the case $\sqrt{K^2}= K$, it corresponds to the triplet dominant case.
The detail algebraic description is given in Ref.\cite{Fukuyama:2015kra}.
On the other hand, for the fit for $c_R v_R \sim 10^{13}$ GeV, which can give a best fit actually,
$f_{11}$ and $f_{12}$ are not small (irrespective of the elements of the neutrino mass matrix), 
and the proton decay amplitudes are not suppressed.

Here, we show concrete examples of the $f$ matrix (in GeV unit) obtained in Ref.\cite{Fukuyama:2015kra}
in the form of $r f v_u$
to compare with $M_d = Y_d v_d = r v_u (h+f)$
in the basis where the down-type quark mass matrix is real (positive) diagonal:
\begin{enumerate} 
\item
$c_R v_R = 8.86 \times 10^{16}$ GeV
\begin{equation}
rfv_u
= \left(
 \begin{array}{ccc}
    0.000127 - 0.000015\,i & 0.000250 + 0.000026 \,i & 0.00153 + 0.00565\, i \\ 
    0.000250 + 0.000026\, i & 0.00782 - 0.00141\, i & 0.0188 - 0.0200 \,i \\
    0.00153 + 0.00565\, i & 0.0188 - 0.0200\, i & -0.299 - 0.063\, i
 \end{array}
\right),
\end{equation}

\item
$c_R v_R = 1.19 \times 10^{13}$ GeV (the best fit)
\begin{equation}
rfv_u
= \left(
 \begin{array}{ccc}
    0.00167 - 0.00007\,i & -0.000678 + 0.000152 \,i & -0.00163 - 0.0150\, i \\ 
   -0.000678 + 0.000152\, i & 0.0312 - 0.00120\, i & 0.0437 - 0.0567 \,i \\
    -0.00163 - 0.0150\, i &  0.0437 - 0.0567\, i & -0.0095 - 0.195\, i
 \end{array}
\right).
\end{equation}

\end{enumerate}
Those two $f_{11}$ components are different by 1 digit. Roughly speaking, the lifetime are different by 2 digit,
and the best fit solution does not satisfy the current experimental bound for the squarks mass to be around 2 TeV.
In the solution for $c_R v_R \agt 10^{16}$ GeV, not only the elements are small, but also
the $f_{11}$ and $f_{12}$ elements are correlated to the (1,1) and (1,2) elements of the neutrino mass matrix
and thus they depends on $\theta_{23}$ and $\delta_{\rm PMNS}$ as seen in the previous section.
As a consequence, we find that the suppressed proton lifetime depends on the neutrino oscillation parameters.

\section{Numerical study of the neutrino oscillation parameter dependence of the proton decay}

The (1,1) and (1,2) elements of the neutrino mass matrix are suppressed
along the curve shown in Fig.1.
As explained in the previous section, 
the smallness of the $f_{11}$ and $f_{12}$ is correlated to the (1,1) and (1,2) elements of the neutrino mass matrix
for the case of $v_R \agt 10^{16}$ GeV.
Therefore, we expect that the proton decay can be suppressed along the curve qualitatively.
Numerically,
even if $({\cal M}_\nu)_{11}= ({\cal M}_\nu)_{12}$, the $f_{11}$ and $f_{12}$ elements are not exactly zero 
due to the type I contribution.
Besides, the smallness of $f_{11}$ and $f_{12}$ to suppress the contribution from $C_R$ is needed in the basis where
the quark mass matrices are diagonal.
Thus, the suppression of the proton decay happens 
at a bit different from the curve shown in Fig.1,
and $\delta_{\rm PMNS}$-$\theta_{23}$ dependence depends on the detail fits of the charged fermion masses and mixings.
In this section, we show the numerical calculation of the proton lifetime 
in order to confirm that the qualitative expectation is realized
using the explicit fits given in Ref.\cite{Fukuyama:2015kra}.

In order to show the numerical calculation of the partial lifetime $p\to K\bar\nu$, we will setup in the following ways.
As given in Eqs.(\ref{cl2}), (\ref{cr2}),
the dimension-five operators $C_{L,R}^{ijkl}$ are given by the linear combinations 
of $h_{ij}h_{kl}$, $f_{ij}h_{kl}$, $h_{ij}f_{kl}$ and $f_{ij}f_{kl}$.
We will choose the three coefficients in $C_L$ to cancel the left-handed contribution to the wino-dressed four-fermi
nucleon decay amplitudes, $A_L(p\to K\bar\nu_\tau)$, $A_L(n\to \pi\bar\nu_\tau)$, and $A_L(n\to K\bar\nu_\tau)$.
The two coefficients in $C_R$ are chosen to cancel the
Higgsino-dressed four-fermi decay amplitudes
$A_R(n\to \pi\bar\nu_\tau)$,and $A_R(n\to K\bar\nu_\tau)$.
As we stressed, the coefficients of $f_{ij}h_{kl}$ have opposite signs in $C_L$ and $C_R$.
Therefore, the coefficients are all fixed by the above conditions up to the overall factor,
and the dominant contribution to $p\to K\nu$ is
the Higgsino-dressed four-four fermi operator $(u d)_R (s \nu_\tau)_L$
and $(us)_R (d\nu_\tau)_L$, as explained in the previous section.
We note that this method does not provide the minimization of the partial decay width.
The partial amplitudes are not necessarily canceled exactly to minimize the decay width.
For example, there are additional freedom to cancel the
amplitudes between the wino-dressed four-fermi operator  $(ud)_L (s\nu_\tau)_L$ 
and the Higgsino-dressed operator $(ud)_R (s\nu_\tau)_L$,
by choosing a phase in the Yukawa coupling, the wino and Higgsino masses.
Our purpose to use this method
is to show
that the qualitative picture given in the previous section appears in the 
numerical calculation.

For the triplet Higgs mass, we suppose
$M_{T_1} \simeq 2 \times 10^{16}$ GeV.
Because the triplet Higgs mixings are multiplied,
$(M_T^{-1})_{11} \simeq X_{11} Y_{11}/M_{T_1}$ depends on the detail of the triplet Higgs mass matrix.
As shown in section 2, the freedom of the bottom Yukawa coupling (namely, the freedom of $\tan\beta$ in MSSM)
depends on $r_1 = U_{11}/V_{11}$.
Naively, the $\tan\beta$ freedom comes from the couplings $H\Delta \Phi$ and $H \overline\Delta \Phi$,
where $\Phi$ is a $\bf 210$ Higgs representation,
and $Y_{11}$ can relates to the $\tan\beta$ freedom.
Therefore, to show the numerical results,
we choose
$(M_T^{-1})_{11} = m_b \tan\beta/m_t/(2 \times 10^{16}) \ {\rm GeV}^{-1}$.
In the fit in Ref.\cite{Fukuyama:2015kra}, we use $\tan\beta=10$.
The Higgsino-dressing amplitude of $p\to K\nu_\tau$ depends on $\tau$ Yukawa coupling, 
and thus the decay amplitude is roughly proportional to $\tan^2\beta$
and the partial lifetime is proportional to $1/\tan^4\beta$ under these setups.

The squark masses and Higgsino mass $\mu$ is also important to calculate the
proton decay amplitudes.
We choose $m_{\tilde q} = \mu = 2$ TeV.
We note that the amplitude is roughly proportional to $\mu/m_{\tilde q}^2$.
For the parameters of hadron matrix elements, we use numerical values given in Ref.\cite{Goto:1998qg}.

\begin{figure}[t]
\begin{center}
\includegraphics[width=0.5\textwidth]{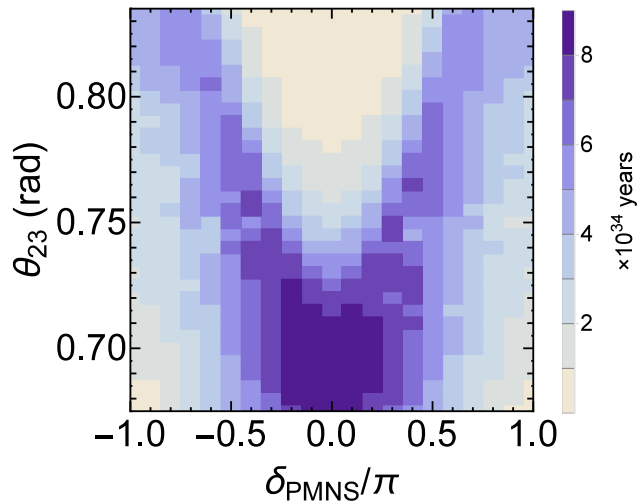}
\caption{
We show the numerical result of the partial lifetime of the $p\to K\bar\nu$ decay.
The detail setup is given in the text.
As expected, the lifetime can be larger near the curve given in Fig.1.
The current experimental bound is
$\tau(p\to K\bar\nu)^{\rm EXP} > 0.59 \times 10^{34}$ years \cite{Abe:2014mwa}.
}
\end{center}
\label{run1}
\end{figure}

In Fig.2, we show the plot of $\tau(p\to K\bar \nu)$ in the $\delta_{\rm PMNS}$-$\theta_{23}$ plane.
As expected, the partial proton lifetime is larger near the curve given in Fig.1.
Near the curve, the lifetime is about 10 times bigger than the current experimental bound
$\tau(p\to K\bar\nu) > 0.59 \times 10^{34}$ years \cite{Abe:2014mwa}.
As described, the lifetime depends on the squark and Higgsino masses $\mu$ and $\tan\beta$.
In the calculation, we use $m_{\tilde q} = \mu = 2$ TeV, and $\tan\beta=10$.
For $\tan\beta = 20$, the lifetime along the curve can reach the current bound for the same SUSY masses.
The accurate measurements of the $\delta_{\rm PMNS}$ phase and the 23-mixing will be very important
if the SUSY particles are found at the LHC.

\section{Conclusions}

The accurate measurement of the PMNS phase is important 
not only to accumulate the fundamental parameter in natural science,
but also to determine the hierarchical structure of the neutrino mass matrix.
If the SUSY particles are found at the LHC, the structure can open the window
of the unified picture of the flavor physics.
The accurate measurement of the neutrino 23-mixing angle is also important 
to obtain the structure.
For the current experimental situation, the deviation of the maximal mixing $\theta_{23} = \pi/4$ is recently reported \cite{nova},
and both $\theta_{23}>\pi/4$ and $\theta_{23}<\pi/4$ are possible within the statistical errors.
It is expected that their accurate measurements can be done in the near future at Tokai-to-HyperKamiokande \cite{Abe:2014oxa}
and DUNE experiments \cite{Acciarri:2016crz}.
The facilities can also raise the bounds of the baryon number violating nucleon decays.
We learned that those two physics can be related in the $SO(10)$ model,
and the accurate measurements of the oscillation parameters are important to
calculate proton decay amplitudes.

The proton decay amplitudes are suppressed at a curve given in Fig.1
if the VEV $(v_R)$ of $\overline{\bf126}$ Higgs representation, which reduces the rank of the $SO(10)$ gauge group,
is more than about $10^{16}$ GeV.
To realize this qualitative behavior,
 the size of the (1,1) and (1,2) elements needs to be correlated between the
$\overline{\bf126}$ coupling matrix $f$ and the neutrino mass matrix.
This algebraic reason is given in the minimal $SO(10)$ model in which the
fermions couple to only $\bf 10$+$\overline{\bf126}$ Higgs fields in Ref.\cite{Fukuyama:2015kra}.
We expect that this qualitative behavior can be realized even in the extended models,
for example, the model that the fermions can also couple with ${\bf 120}$ Higgs field,
since it is due to the algebraic feature. 
The concrete quantitative numbers can be surely different due to the fit of the fermion masses and mixings.
The minimal $SO(10)$ model is instructive to learn how this qualitative behavior of the suppressed proton decay is obtained.
The important ingredient is $v_R \agt 10^{16}$ GeV.
As noted in Ref.\cite{Fukuyama:2015kra},
the influence of $v_R \ll 10^{16}$ GeV to low energy physics 
is the induction of flavor changing neutral currents in the right-handed charged sleptons,
which can enlarge the electron electric dipole moment,
and we can construct a big picture of the unified framework by the future experimental results.

%%%%%%%%%%%%%%%%%%%%%%%%%%%%%%%%%%%%%%%%%%%%%%%%%%%%%%%%%%%%%%%%%%%%%%%%%%

\section*{Acknowledgments}

The work of T.F.\ is supported in part by the Grant-in-Aid for
Science Research from the Ministry of Education, Science and
Culture of Japan (No.\ 26247036).
The work of K.I. is supported in part by JSPS Research Fellowships for Young Scientists. 
The work of Y.M. is supported by the Excellent Research Projects of
 National Taiwan University under grant number NTU-ERP-105R8915. 

%%%%%%%%%%%%%%%%%%%%%%%%%%%%%%%%%%%%%%%%%%%%%%%%%%%%%%%%%%%%%%%%%%%%%%%%%


\begin{thebibliography}{99}






 \bibitem{seesaw} 
 P.~Minkowski,
%{\it {mu $\to$ e gamma at a Rate of One Out of
%1-Billion Muon  Decays?}},
Phys. Lett. {\bf B67}, 421 (1977);
%%CITATION = PHLTA,B67,421;%%
%
%\bibitem{Yanagida}
T.~Yanagida in {\em Workshop on Unified Theories, KEK Report 79-18}, p.~95 (1979);
%
%\bibitem{Gell-Mann}
M.~Gell-Mann, P.~Ramond and R.~Slansky, {\em Supergravity},
p.~315; Amsterdam: North Holland (1979);
%
%\bibitem{Glashow}
S.~L. Glashow, {\em 1979 Cargese Summer Institute on Quarks and
Leptons}, p.~687; New York: Plenum (1980);
%
%\bibitem{Mohapatra:1980yp}
R.~N. Mohapatra and G.~Senjanovic,
%{\it {Neutrino Masses and Mixings in Gauge
%  Models with Spontaneous Parity Violation}},
Phys. Rev. Lett. {\bf 44}, 912 (1980).


\bibitem{Schechter:1980gr}
 %
%\cite{Lazarides:1980nt}
    J.~Schechter and J.~W.~F.~Valle,
  %``Neutrino Masses in SU(2) x U(1) Theories,''
  Phys.\ Rev.\ D {\bf 22}, 2227 (1980);
%
%\bibitem{Lazarides:1980nt}
  G.~Lazarides, Q.~Shafi and C.~Wetterich,
  %``Proton Lifetime And Fermion Masses In An SO(10) Model,''
  Nucl.\ Phys.\ B {\bf 181}, 287 (1981);
  %%CITATION = NUPHA,B181,287;%
  %
   R.~N.~Mohapatra and G.~Senjanovic,
  %``Neutrino Masses And Mixings In Gauge Models With Spontaneous Parity
  %Violation,''
  Phys.\ Rev.\ D {\bf 23}, 165 (1981).





%\cite{An:2012eh}
\bibitem{An:2012eh} 
  F.~P.~An {\it et al.}  [DAYA-BAY Collaboration],
  %``Observation of electron-antineutrino disappearance at Daya Bay,''
  Phys.\ Rev.\ Lett.\  {\bf 108}, 171803 (2012)
  [arXiv:1203.1669 [hep-ex]];
  %%CITATION = ARXIV:1203.1669;%%
%
%\cite{Ahn:2012nd}
%\bibitem{Ahn:2012nd} 
  J.~K.~Ahn {\it et al.}  [RENO Collaboration],
  %``Observation of Reactor Electron Antineutrino Disappearance in the RENO Experiment,''
  Phys.\ Rev.\ Lett.\  {\bf 108}, 191802 (2012)
  [arXiv:1204.0626 [hep-ex]];
  %%CITATION = ARXIV:1204.0626;%%
%
%\cite{Abe:2011sj}
%\bibitem{Abe:2011sj} 
  K.~Abe {\it et al.}  [T2K Collaboration],
  %``Indication of Electron Neutrino Appearance from an Accelerator-produced Off-axis Muon Neutrino Beam,''
  Phys.\ Rev.\ Lett.\  {\bf 107}, 041801 (2011)
  [arXiv:1106.2822 [hep-ex]].
  %%CITATION = ARXIV:1106.2822;%%

%\cite{Wilking:2013vza}
\bibitem{Wilking:2013vza} 
%
%\cite{Adamson:2013ue}
%\bibitem{Adamson:2013ue} 
  P.~Adamson {\it et al.} [MINOS Collaboration],
  %``Electron neutrino and antineutrino appearance in the full MINOS data sample,''
  Phys.\ Rev.\ Lett.\  {\bf 110}, no. 17, 171801 (2013)
  [arXiv:1301.4581 [hep-ex]];
  %%CITATION = ARXIV:1301.4581;%%
  %94 citations counted in INSPIRE as of 29 juil. 2015
%
%\cite{Wilking:2013vza}
%\bibitem{Wilking:2013vza} 
  M.~Wilking [T2K Collaboration],
  %``New Results from the T2K Experiment: Observation of $\nu_e$ Appearance in a $\nu_\mu$ Beam,''
  PoS EPS {\bf -HEP2013}, 536 (2013)
  [arXiv:1311.4114 [hep-ex]];
  %%CITATION = ARXIV:1311.4114;%%
  %6 citations counted in INSPIRE as of 10 juin 2015
%
%\cite{Adamson:2016tbq}
%\bibitem{Adamson:2016tbq} 
  P.~Adamson {\it et al.} [NOvA Collaboration],
  %``First measurement of electron neutrino appearance in NOvA,''
  Phys.\ Rev.\ Lett.\  {\bf 116}, no. 15, 151806 (2016)
%  doi:10.1103/PhysRevLett.116.151806
  [arXiv:1601.05022 [hep-ex]];
  %%CITATION = doi:10.1103/PhysRevLett.116.151806;%%
  %14 citations counted in INSPIRE as of 05 May 2016
% 
%\cite{Adamson:2016xxw}
%\bibitem{Adamson:2016xxw} 
  P.~Adamson {\it et al.} [NOvA Collaboration],
  %``First measurement of muon-neutrino disappearance in NOvA,''
  Phys.\ Rev.\ D {\bf 93}, no. 5, 051104 (2016)
%  doi:10.1103/PhysRevD.93.051104
  [arXiv:1601.05037 [hep-ex]].
  %%CITATION = doi:10.1103/PhysRevD.93.051104;%%
  %9 citations counted in INSPIRE as of 05 May 2016
%\cite{Adamson:2016xxw}


%\cite{GonzalezGarcia:2012sz}
\bibitem{GonzalezGarcia:2012sz} 
  M.~C.~Gonzalez-Garcia, M.~Maltoni, J.~Salvado and T.~Schwetz,
  %``Global fit to three neutrino mixing: critical look at present precision,''
  JHEP {\bf 1212}, 123 (2012)
  [arXiv:1209.3023 [hep-ph]];
  %%CITATION = ARXIV:1209.3023;%%
  %488 citations counted in INSPIRE as of 29 juil. 2015
%\cite{Forero:2014bxa}
%\bibitem{Forero:2014bxa} 
  D.~V.~Forero, M.~Tortola and J.~W.~F.~Valle,
  %``Neutrino oscillations refitted,''
  Phys.\ Rev.\ D {\bf 90}, no. 9, 093006 (2014)
  [arXiv:1405.7540 [hep-ph]].
  %%CITATION = ARXIV:1405.7540;%%
  %149 citations counted in INSPIRE as of 29 Jul 2015

%\cite{Capozzi:2013csa}
\bibitem{Capozzi:2013csa} 
  F.~Capozzi, G.~L.~Fogli, E.~Lisi, A.~Marrone, D.~Montanino and A.~Palazzo,
  %``Status of three-neutrino oscillation parameters, circa 2013,''
  Phys.\ Rev.\ D {\bf 89}, 093018 (2014)
  [arXiv:1312.2878 [hep-ph]];
  %%CITATION = ARXIV:1312.2878;%%
  %192 citations counted in INSPIRE as of 04 Jun 2015
%
%\cite{Capozzi:2016rtj}
%\bibitem{Capozzi:2016rtj} 
  F.~Capozzi, E.~Lisi, A.~Marrone, D.~Montanino and A.~Palazzo,
  %``Neutrino masses and mixings: Status of known and unknown $3\nu$ parameters,''
  Nucl.\ Phys.\ B {\bf 908}, 218 (2016)
  %  doi:10.1016/j.nuclphysb.2016.02.016
  [arXiv:1601.07777 [hep-ph]].
  %%CITATION = doi:10.1016/j.nuclphysb.2016.02.016;%%
  %38 citations counted in INSPIRE as of 26 Sep 2016






%\cite{Dutta:2009ij}
\bibitem{Dutta:2009ij} 
  B.~Dutta, Y.~Mimura and R.~N.~Mohapatra,
  %``Origin of Quark-Lepton Flavor in SO(10) with Type II Seesaw,''
  Phys.\ Rev.\ D {\bf 80}, 095021 (2009)
  [arXiv:0910.1043 [hep-ph]];
  %%CITATION = ARXIV:0910.1043;%%
  %35 citations counted in INSPIRE as of 12 Jun 2015
%\cite{Dutta:2009bj}
%\bibitem{Dutta:2009bj} 
%  B.~Dutta, Y.~Mimura and R.~N.~Mohapatra,
  %``An SO(10) Grand Unified Theory of Flavor,''
  JHEP {\bf 1005}, 034 (2010)
  [arXiv:0911.2242 [hep-ph]]; 
  %%CITATION = ARXIV:0911.2242;%%
  %58 citations counted in INSPIRE as of 12 juin 2015
%\cite{BhupalDev:2012nm}
%\bibitem{BhupalDev:2012nm} 
  P.~S.~Bhupal Dev, B.~Dutta, R.~N.~Mohapatra and M.~Severson,
  %``$\theta_{13}$ and Proton Decay in a Minimal $SO(10) \times S_4$ model of Flavor,''
  Phys.\ Rev.\ D {\bf 86}, 035002 (2012)
  doi:10.1103/PhysRevD.86.035002
  [arXiv:1202.4012 [hep-ph]].
  %%CITATION = doi:10.1103/PhysRevD.86.035002;%%
  %47 citations counted in INSPIRE as of 30 Sep 2016



%\cite{Weinberg:1981wj}
\bibitem{Weinberg:1981wj} 
  S.~Weinberg,
  %``Supersymmetry at Ordinary Energies. 1. Masses and Conservation Laws,''
  Phys.\ Rev.\ D {\bf 26}, 287 (1982);
%  doi:10.1103/PhysRevD.26.287
  %%CITATION = doi:10.1103/PhysRevD.26.287;%%
  %1000 citations counted in INSPIRE as of 12 Sep 2016
%  
%\cite{Sakai:1981pk}
%\bibitem{Sakai:1981pk} 
  N.~Sakai and T.~Yanagida,
  %``Proton Decay in a Class of Supersymmetric Grand Unified Models,''
  Nucl.\ Phys.\ B {\bf 197}, 533 (1982).
%  doi:10.1016/0550-3213(82)90457-6
  %%CITATION = doi:10.1016/0550-3213(82)90457-6;%%
  %816 citations counted in INSPIRE as of 12 Sep 2016  
  




\bibitem{Dutta:2004zh} 
  B.~Dutta, Y.~Mimura and R.~N.~Mohapatra,
  %``Suppressing proton decay in the minimal SO(10) model,''
  Phys.\ Rev.\ Lett.\  {\bf 94}, 091804 (2005)
  [hep-ph/0412105];
  %%CITATION = HEP-PH/0412105;%%
%
%\cite{Dutta:2005ni}
%\bibitem{Dutta:2005ni} 
%  B.~Dutta, Y.~Mimura and R.~N.~Mohapatra,
  %``Neutrino mixing predictions of a minimal SO(10) model with suppressed proton decay,''
  Phys.\ Rev.\ D {\bf 72}, 075009 (2005)
  [hep-ph/0507319].
  %%CITATION = HEP-PH/0507319;%%
 




  
%\cite{Fukuyama:2015kra}
\bibitem{Fukuyama:2015kra} 
  T.~Fukuyama, K.~Ichikawa and Y.~Mimura,
  %``Revisiting fermion mass and mixing fits in the minimal SUSY $SO(10)$ GUT,''
  Phys.\ Rev.\ D {\bf 94}, no. 7, 075018 (2016)
  doi:10.1103/PhysRevD.94.075018
  [arXiv:1508.07078 [hep-ph]].
  %%CITATION = doi:10.1103/PhysRevD.94.075018;%%
  %6 citations counted in INSPIRE as of 18 Nov 2016  







\bibitem{Babu:1992ia}
  K.~S.~Babu and R.~N.~Mohapatra,
  %``Predictive Neutrino Spectrum In Minimal SO(10) Grand Unification,''
  Phys.\ Rev.\ Lett.\  {\bf 70}, 2845 (1993)
  [hep-ph/9209215].





 %\cite{Matsuda:2000zp}
\bibitem{Matsuda:2000zp} 
  K.~Matsuda, Y.~Koide and T.~Fukuyama,
  %``Can the SO(10) model with two Higgs doublets reproduce the observed fermion masses?,''
  Phys.\ Rev.\ D {\bf 64}, 053015 (2001)
  [hep-ph/0010026];
  %%CITATION = HEP-PH/0010026;%%
%\cite{Matsuda:2001bg}
%
%\bibitem{Matsuda:2001bg} 
  K.~Matsuda, Y.~Koide, T.~Fukuyama and H.~Nishiura,
  %``How far can the SO(10) two Higgs model describe the observed neutrino masses and mixings?,''
  Phys.\ Rev.\ D {\bf 65}, 033008 (2002)
  [Erratum-ibid.\ D {\bf 65}, 079904 (2002)]
  [hep-ph/0108202].
  %%CITATION = HEP-PH/0108202;%%





 \bibitem{Fukuyama:2002ch} 
  T.~Fukuyama and N.~Okada,
  %``Neutrino oscillation data versus minimal supersymmetric SO(10) model,''
  JHEP {\bf 0211}, 011 (2002)
  [hep-ph/0205066].
  %%CITATION = HEP-PH/0205066.%%
%  
 
 
\bibitem{clark}
  T.~E.~Clark, T.~K.~Kuo and N.~Nakagawa,
  %``A SO(10) Supersymmetric Grand Unified Theory,''
  Phys.\ Lett.\  B {\bf 115}, 26 (1982); 
  %
  C.~S.~Aulakh and R.~N.~Mohapatra,
  %``Implications Of Supersymmetric SO(10) Grand Unification,''
  Phys.\ Rev.\  D {\bf 28}, 217 (1983);
%\bibitem{lee}
 % D.~G.~Lee,
  %``Symmetry breaking and mass spectra in the minimal supersymmetric SO(10)
  %grand unified theory,''
 % Phys.\ Rev. {\bf D49}, 1417 (1994).
%
%\bibitem{aulakh}
  C.~S.~Aulakh, B.~Bajc, A.~Melfo, G.~Senjanovi\'c and F.~Vissani,
  %``The minimal supersymmetric grand unified theory,''
  Phys.\ Lett.\  B {\bf 588}, 196 (2004)
  [arXiv:hep-ph/0306242].
 
 
%\bibitem{Fukuyama:2004xs} 
 % T.~Fukuyama, A.~Ilakovac, T.~Kikuchi, S.~Meljanac and N.~Okada,
  %``General formulation for proton decay rate in minimal supersymmetric SO(10) GUT,''
  %Eur.\ Phys.\ J.\ C {\bf 42}, 191 (2005)
 % [hep-ph/0401213]: Phys.\ Rev.\ D {\bf 72}, 051701 (2005)
 % [hep-ph/0412348].


%\cite{Aulakh:2004hm}
\bibitem{Aulakh:2004hm} 
T.~Fukuyama, A.~Ilakovac, T.~Kikuchi, S.~Meljanac and N.~Okada,
  %``General formulation for proton decay rate in minimal supersymmetric SO(10) GUT,''
  Eur.\ Phys.\ J.\ C {\bf 42}, 191 (2005)  [hep-ph/0401213];
%  
%\bibitem{Fukuyama:2004ti} 
%  T.~Fukuyama, A.~Ilakovac, T.~Kikuchi, S.~Meljanac and N.~Okada,
  %``Higgs masses in the minimal SUSY SO(10) GUT,''
  Phys.\ Rev.\ D {\bf 72}, 051701 (2005)
  [hep-ph/0412348];
  %%CITATION = HEP-PH/0412348;%%
%
  B.~Bajc, A.~Melfo, G.~Senjanovic and F.~Vissani,
 %``The Minimal supersymmetric grand unified theory. 1. Symmetry breaking and the particle spectrum,''
 Phys.\ Rev.\ D {\bf 70}, 035007 (2004)
 [hep-ph/0402122]; 
%
%\bibitem{Bajc:2005qe} 
 % B.~Bajc, A.~Melfo, G.~Senjanovic and F.~Vissani,
  %``Fermion mass relations in a supersymmetric SO(10) theory,''
  Phys.\ Lett.\ B {\bf 634}, 272 (2006)
  [hep-ph/0511352];
  %%CITATION = HEP-PH/0511352;%%
 %
  C.~S.~Aulakh and A.~Girdhar,
  %``SO(10) MSGUT: Spectra, couplings and threshold effects,''
  Nucl.\ Phys.\ B {\bf 711}, 275 (2005)
  [hep-ph/0405074];
  %%CITATION = HEP-PH/0405074;%%
  %
  %\cite{Fukuyama:2004ps}
%\bibitem{Fukuyama:2004ps} 
  T.~Fukuyama, A.~Ilakovac, T.~Kikuchi, S.~Meljanac and N.~Okada,
  %``SO(10) group theory for the unified model building,''
  J.\ Math.\ Phys.\  {\bf 46}, 033505 (2005)
  [hep-ph/0405300].
  %%CITATION = HEP-PH/0405300;%%
  %104 citations counted in INSPIRE as of 15 juin 2015
%   
%\cite{Fukuyama:2004ti}
%\bibitem{Fukuyama:2004ti} 
%  T.~Fukuyama, A.~Ilakovac, T.~Kikuchi, S.~Meljanac and N.~Okada,
  %``Higgs masses in the minimal SUSY SO(10) GUT,''
 % Phys.\ Rev.\ D {\bf 72}, 051701 (2005)
  %[hep-ph/0412348].
  %%CITATION = HEP-PH/0412348;%%  
  




 \bibitem{Bajc:2002iw} 
  B.~Bajc, G.~Senjanovic and F.~Vissani,
  %``b - tau unification and large atmospheric mixing: A Case for noncanonical seesaw,''
  Phys.\ Rev.\ Lett.\  {\bf 90}, 051802 (2003)
  [hep-ph/0210207];
  %%CITATION = HEP-PH/0210207;%%
  %
%\cite{Bajc:2004fj}
%\bibitem{Bajc:2004fj} 
%  B.~Bajc, G.~Senjanovic and F.~Vissani,
  %``Probing the nature of the seesaw in renormalizable SO(10),''
  Phys.\ Rev.\ D {\bf 70}, 093002 (2004)
  [hep-ph/0402140];
  %%CITATION = HEP-PH/0402140;%%  
%  
%
%    \bibitem{Goh:2003sy} 
 H.~S.~Goh, R.~N.~Mohapatra and S.~-P.~Ng,
  %``Minimal SUSY SO(10), b tau unification and large neutrino mixings,''
 Phys.\ Lett.\ B {\bf 570}, 215 (2003)
  [hep-ph/0303055];
  %
%\cite{Goh:2003hf}
%\bibitem{Goh:2003hf} 
% H.~S.~Goh, R.~N.~Mohapatra and S.~-P.~Ng,
  %``Minimal SUSY SO(10) model and predictions for neutrino mixings and leptonic CP violation,''
  Phys.\ Rev.\ D {\bf 68}, 115008 (2003)
  [hep-ph/0308197];
  %%CITATION = HEP-PH/0308197;%%
%
%
%\bibitem{Dutta:2004wv}   
  %``CKM CP violation in a minimal SO(10) model for neutrinos and its implications,''
    %%CITATION = HEP-PH/0406262;%%
%
%
%\bibitem{Dutta:2004wv}
 B.~Dutta, Y.~Mimura and R.~N.~Mohapatra,
Phys.\ Rev.\ D {\bf 69}, 115014 (2004)
  [hep-ph/0402113];
  %%CITATION = HEP-PH/0402113;%%
%
%\cite{Dutta:2004hp}
%\bibitem{Dutta:2004hp} 
%  B.~Dutta, Y.~Mimura and R.~N.~Mohapatra,
  %``Neutrino masses and mixings in a predictive SO(10) model with CKM CP violation,''
  Phys.\ Lett.\ B {\bf 603}, 35 (2004)
  [hep-ph/0406262].
 %%CITATION = PRLTA,100,181801;%%
 %





  

%\cite{Babu:2005ia}
%
%\cite{Bertolini:2006pe}
 %  G.~Altarelli and G.~Blankenburg,
  %``Different $SO(10)$ Paths to Fermion Masses and Mixings,''
  %JHEP {\bf 1103}, 133 (2011)
 % [arXiv:1012.2697 [hep-ph]];
 %
  %%CITATION = HEP-PH/0605006;%%
 % A.~S.~Joshipura and K.~M.~Patel,
 % %``Fermion Masses in SO(10) Models,''
 % Phys.\ Rev.\ D {\bf 83}, 095002 (2011)
 % [arXiv:1102.5148 [hep-ph]]:
  %%CITATION = HEP-PH/0406117;%%
  %
 

\bibitem{Babu:2005ia}
%\cite{Bertolini:2006pe}
    K.~S.~Babu and C.~Macesanu,
  %``Neutrino masses and mixings in a minimal SO(10) model,''
  Phys.\ Rev.\ D {\bf 72}, 115003 (2005)
  [hep-ph/0505200].
  %%CITATION = HEP-PH/0505200;%%



 \bibitem{Bertolini:2006pe} 
  S.~Bertolini, T.~Schwetz and M.~Malinsky,
  %``Fermion masses and mixings in SO(10) models and the neutrino challenge to SUSY GUTs,''
  Phys.\ Rev.\ D {\bf 73}, 115012 (2006)
  [hep-ph/0605006].







%\cite{Agashe:2014kda}
\bibitem{Agashe:2014kda} 
  K.~A.~Olive {\it et al.}  [Particle Data Group Collaboration],
  %``Review of Particle Physics,''
  Chin.\ Phys.\ C {\bf 38}, 090001 (2014).
  %%CITATION = CHPHD,C38,090001;%%
  %1317 citations counted in INSPIRE as of 26 juin 2015
  


%\cite{Fritzsch:2011qv}
\bibitem{Fritzsch:2011qv} 
  H.~Fritzsch, Z.~z.~Xing and S.~Zhou,
  %``Two-zero Textures of the Majorana Neutrino Mass Matrix and Current Experimental Tests,''
  JHEP {\bf 1109}, 083 (2011)
  [arXiv:1108.4534 [hep-ph]].
  %%CITATION = ARXIV:1108.4534;%%
  %78 citations counted in INSPIRE as of 12 Aug 2015


%\cite{Geng:2015oga}
\bibitem{Geng:2015oga} 
  C.~Q.~Geng, D.~Huang and L.~H.~Tsai,
  %``CP violations in predictive neutrino mass structures,''
  Eur.\ Phys.\ J.\ C {\bf 75}, no. 11, 557 (2015)
%  doi:10.1140/epjc/s10052-015-3779-9
  [arXiv:1508.02180 [hep-ph]].
  %%CITATION = doi:10.1140/epjc/s10052-015-3779-9;%%




%\cite{Haba:2011pe}
\bibitem{Haba:2011pe} 
  N.~Haba, T.~Horita, K.~Kaneta and Y.~Mimura,
  %``TeV-scale seesaw with non-negligible left-right neutrino mixings,''
  arXiv:1110.2252 [hep-ph].
  %%CITATION = ARXIV:1110.2252;%%
  %9 citations counted in INSPIRE as of 12 Aug 2015
  

%\cite{Dutta:2013bvf}
\bibitem{Dutta:2013bvf} 
  B.~Dutta, Y.~Mimura and R.~N.~Mohapatra,
  %``Proton decay and $\mu\to e+\gamma$ connection in a renormalizable SO(10) GUT for neutrinos,''
  Phys.\ Rev.\ D {\bf 87}, no. 7, 075008 (2013)
  [arXiv:1302.2574 [hep-ph]].
  %%CITATION = ARXIV:1302.2574;%%
  %1 citations counted in INSPIRE as of 12 Jun 2015





%\cite{Cheng:1980qt}
\bibitem{Cheng:1980qt} 
  T.~P.~Cheng and L.~F.~Li,
  %``Neutrino Masses, Mixings and Oscillations in SU(2) x U(1) Models of Electroweak Interactions,''
  Phys.\ Rev.\ D {\bf 22}, 2860 (1980).
  %%CITATION = PHRVA,D22,2860;%%








%\cite{Goto:1998qg}
\bibitem{Goto:1998qg} 
  T.~Goto and T.~Nihei,
  %``Effect of RRRR dimension five operator on the proton decay in the minimal SU(5) SUGRA GUT model,''
  Phys.\ Rev.\ D {\bf 59}, 115009 (1999)
  [hep-ph/9808255].
  %%CITATION = HEP-PH/9808255;%%
  %202 citations counted in INSPIRE as of 04 juin 2015

%\cite{Murayama:2001ur}
\bibitem{Murayama:2001ur} 
  H.~Murayama and A.~Pierce,
  %``Not even decoupling can save minimal supersymmetric SU(5),''
  Phys.\ Rev.\ D {\bf 65}, 055009 (2002)
  [hep-ph/0108104].
  %%CITATION = HEP-PH/0108104;%%
  %300 citations counted in INSPIRE as of 15 juin 2015






%\cite{Goh:2003nv}
\bibitem{Goh:2003nv} 
  H.~S.~Goh, R.~N.~Mohapatra, S.~Nasri and S.~P.~Ng,
  %``Proton decay in a minimal SUSY SO(10) model for neutrino mixings,''
  Phys.\ Lett.\ B {\bf 587}, 105 (2004)
  [hep-ph/0311330];
  %%CITATION = HEP-PH/0311330;%%
%
%\cite{Fukuyama:2004xs}
%\bibitem{Fukuyama:2004xs} 
 % T.~Fukuyama, A.~Ilakovac, T.~Kikuchi, S.~Meljanac and N.~Okada,
 %``General formulation for proton decay rate in minimal supersymmetric SO(10) GUT,''
 % Eur.\ Phys.\ J.\ C {\bf 42}, 191 (2005)
 % [hep-ph/0401213];
  %%CITATION = HEP-PH/0401213;%%
% 
 %\cite{Fukuyama:2004pb}
%\bibitem{Fukuyama:2004pb} 
  T.~Fukuyama, A.~Ilakovac, T.~Kikuchi, S.~Meljanac and N.~Okada,
  %``Detailed analysis of proton decay rate in the minimal supersymmetric SO(10) model,''
  JHEP {\bf 0409}, 052 (2004)
  [hep-ph/0406068];
  %%CITATION = HEP-PH/0406068;%%
%
%\cite{Severson:2015dta}
%\bibitem{Severson:2015dta} 
  M.~Severson,
  %``Neutrino Sector and Proton Lifetime in a Realistic SUSY SO(10) Model,''
  Phys.\ Rev.\ D {\bf 92}, no. 9, 095026 (2015)
  [arXiv:1506.08468 [hep-ph]];
  %%CITATION = doi:10.1103/PhysRevD.92.095026;%%
%
%\cite{Severson:2016rqb}
%\bibitem{Severson:2016rqb} 
%  M.~Severson,
  %``Neutrino Mass and Proton Decay in a Realistic Supersymmetric SO(10) Model,''
  arXiv:1601.06478 [hep-ph].
  %%CITATION = ARXIV:1601.06478;%%




%\cite{Abe:2014mwa}
\bibitem{Abe:2014mwa} 
  K.~Abe {\it et al.} [Super-Kamiokande Collaboration],
  %``Search for proton decay via $p\to\nu��K^+$ using 260??kiloton?year data of Super-Kamiokande,''
  Phys.\ Rev.\ D {\bf 90}, no. 7, 072005 (2014)
  doi:10.1103/PhysRevD.90.072005
  [arXiv:1408.1195 [hep-ex]].
  %%CITATION = doi:10.1103/PhysRevD.90.072005;%%
  %50 citations counted in INSPIRE as of 12 Sep 2016



\bibitem{nova}
 Talk at Neutrino conference in London, July 2016, [NOvA Collaboration].
 

%\cite{Abe:2014oxa}
\bibitem{Abe:2014oxa} 
  K.~Abe {\it et al.} [Hyper-Kamiokande Working Group Collaboration],
  %``A Long Baseline Neutrino Oscillation Experiment Using J-PARC Neutrino Beam and Hyper-Kamiokande,''
  arXiv:1412.4673 [physics.ins-det].
  %%CITATION = ARXIV:1412.4673;%%
  %33 citations counted in INSPIRE as of 12 Sep 2016
    
  
  
%\cite{Acciarri:2016crz}
\bibitem{Acciarri:2016crz} 
  R.~Acciarri {\it et al.} [DUNE Collaboration],
  %``Long-Baseline Neutrino Facility (LBNF) and Deep Underground Neutrino Experiment (DUNE) : Volume 1: The LBNF and DUNE Projects,''
  arXiv:1601.05471 [physics.ins-det].
  %%CITATION = ARXIV:1601.05471;%%
  %33 citations counted in INSPIRE as of 12 Sep 2016
  
    

\end{thebibliography}
\end{document}